\documentclass[twocolumn]{raa}
 
\usepackage{graphicx,times}             
\usepackage{natbib}
\usepackage{amssymb,amsmath}
\usepackage{mathrsfs}
\bibpunct{(}{)}{;}{a}{}{,}

\usepackage[pagebackref=true]{hyperref}

\newcommand{\mpc}{{\, {\rm Mpc}}}

\def\aj{AJ}

\def\apjs{ApJS}

\def\mnras{MNRAS}
\def\aap{A\&A}

\def\nat{Nature}      
\def\apjs{ApJS}
\def\apjl{ApJ Letters}

\begin{document}

\title{Galaxy interactions in filaments and sheets: effects of the large-scale structures versus the local density}

   \volnopage{Vol.0 (20xx) No.0, 000--000}     
   \setcounter{page}{1}           

   \author{Apashanka Das
      \inst{1}
   \and Biswajit Pandey
      \inst{2}
   \and Suman Sarkar
      \inst{3}
   }
 \institute{Department of Physics, Visva-Bharati University, Santiniketan, 
              Birbhum, 731235, India
             {\it a.das.cosmo@gmail.com}\\
        \and
	     {Department of Physics, Visva-Bharati University, Santiniketan, 
	     Birbhum, 731235, India {\it biswap@visva-bharati.ac.in}}\\
        \and
            {Department of Physics, Indian Institute of Science Education and
              Research Tirupati, Tirupati - 517507, Andhra Pradesh, India
              {\it suman2reach@gmail.com}}\\
   }
\vs\no
   {\small Received 20xx month day; accepted 20xx month day}

 \abstract{The major interactions are known to trigger star formation
   in galaxies and alter their colour. We study the major interactions
   in filaments and sheets using the SDSS data to understand the
   influence of large-scale environments on the galaxy
   interactions. We identify the galaxies in filaments and sheets
   using the local dimension and also find the major pairs residing in
   these environments. The star formation rate and colour of the
   interacting galaxies as a function of pair separation are
   separately analyzed in filaments and sheets. The analysis is
   repeated for three volume limited samples covering different
   magnitude ranges. The major pairs residing in the filaments show a
   significantly higher star formation rate (SFR) and bluer colour
   than those residing in the sheets up to the projected pair
   separation of $\sim 50$ kpc. We observe a complete reversal of this
   behaviour for both the SFR and colour of the galaxy pairs having a
   projected separation larger than 50 kpc. Some earlier studies
   report that the galaxy pairs align with the filament axis. Such
   alignment inside filaments indicates anisotropic accretion that may
   cause these differences. We do not observe these trends in the
   brighter galaxy samples. The pairs in filaments and sheets from the
   brighter galaxy samples trace relatively denser regions in these
   environments. The absence of these trends in the brighter samples
   may be explained by the dominant effect of the local density over
   the effects of the large-scale environment.  }

\authorrunning{A. Das, B. Pandey \& S. Sarkar}
%author_head in even pages
\titlerunning{Galaxy interactions in cosmic web}% title_head in odd pages
 
\maketitle

\section{Introduction}           %% first-level sections will be auto-capitalized
\label{sect:intro}

The present-day Universe is populated with myriad galaxies that are
vast collections of star, gas, dust and dark matter. The galaxies are
the fundamental units of the large-scale structures in the
Universe. The early redshift surveys during the late seventies and
early eighties demonstrate that the galaxies are distributed in a
complex interconnected network surrounded by large empty regions
\citep{gregory78,joeveer78,einasto80,zeldovich82,einasto84}. The
existence of this network of filaments, sheets and clusters encircled
by numerous voids become more evident with the advent of modern galaxy
redshift surveys \citep{stout02, colless01}. The role of the different
geometric environments of the cosmic web \citep{bond96} on galaxy
formation and evolution remains an active area of research since then.

 The galaxies are believed to have formed via the cooling and
 condensation of the accreted neutral hydrogen gas at the centre of
 the dark matter halos \citep{reesostriker77, silk77, white78,
   fall80}. The dark matter halos reside in different morphological
 environments of the cosmic web. Studies with the hydrodynamical
 simulations suggest that the filaments are dominated by gas in WHIM
 that accounts for more than $80\%$ of the baryonic budget in the
 Universe \citep{tuominen21, galarraga21}. It has been suggested by a
 number of works that the filaments play a significant role in
 governing the gas accretion efficiency in the galaxies
 \citep{cornu18, zhu22}. The dark matter halos residing in filaments
 and sheets may have different gas accretion efficiency. An earlier
 analysis shows that the star forming blue galaxies have a more
 filamentary distribution than their red counterparts
 \citep{pandey08}. The large-scale coherent patterns like sheets and
 filaments may play significant roles in the formation and evolution
 of galaxies.

The roles of environment on the formation and evolution of galaxies
have been extensively studied in the literature \citep{oemler74,
  davis76, dress80, guzo97, zevi02, hog03, blan03, einas03, gotto03,
  kauffmann04, pandey06, park07, mocine07, pandey08, porter08,
  bamford09, cooper10, koyama13, pandey17, sarkar20, bhattacharjee20,
  pandey20}. The galaxies interact with their environment and other
galaxies in their neighbourhood. It is well known that the galaxies in
the high density regions have a lower star formation activity
\citep{lewis02,gomez03,kauffmann04}. The quenching of star formation
in high density regions can be induced by a host of mechanisms such as
ram pressure stripping \citep{gunn72}, galaxy harassment
\citep{moore96, moore98}, strangulation \citep{gunn72, balogh00},
starvation \citep{larson80, somerville99, kawata08} and gas loss
through starburst, AGN or shock-driven winds \citep{cox04, murray05,
  springel05}. A galaxy can also quench its star formation through
different physical processes such as mass quenching \citep{birnboim03,
  dekel06, keres05, gabor10}, morphological quenching
\citep{martig09}, bar quenching \citep{masters10} and angular momentum
quenching \citep{peng20}. The galaxy interactions on the other hand
can trigger star formation activity in galaxies and alter their colour
\citep{barton00, lambas03, alonso04, nikolic04, alonso06, woods06,
  woods07, barton07, ellison08, heiderman09, knapen09, robaina09,
  ellison10, woods10, patton11}.

The density of the local environment is known to play a crucial role
in deciding the galaxy properties and their evolution. However, the
roles of the different morphological environments of the cosmic web on
the formation and evolution of galaxies are less clearly
understood. The sheets and filaments provide unique environments for
galaxy formation and evolution. The different physical mechanisms
triggering or quenching star formation in galaxies may be impacted
differently in such environments. In this work, we consider the major
interaction between galaxies in sheets and filaments. The major
interaction between galaxies are known to trigger new star
formation. The galaxy pairs are frequently observed in the denser
regions. Both filaments and sheets represent overdense regions of the
cosmic web and are expected to host a significant number of major
galaxy pairs. The SFR of a galaxy is largely set by the available gas
mass, which itself is modulated by inflows and outflows of gas
\citep{dekel09, dave11, dave12, lilly13}.  The interaction and mergers
are transient events that can push galaxies out of equilibrium. The
differences in the availability of gas and the accretion efficiency of
the interacting galaxies in filaments and sheets may influence their
physical properties.

This work aims to study the differences in the major galaxy
interaction observed in sheets and filaments. Currently, SDSS
\citep{stout02} is the largest redshift survey with the reliable
photometric and spectroscopic information of millions of galaxies in
the nearby Universe. It provides us the unique opportunity to address
such questions in a statistical manner. We construct a set of volume
limited sample of galaxies in different luminosity range. We use the
local dimension \citep{sarkar09} to identify the galaxies residing in
sheets and filaments in the cosmic web. We then find the galaxy pairs
residing in these environments and study their SFR and colour as a
function of the projected pair separation.

We use both SFR and colour of the galaxies in major pairs for the
present analysis. The enhancement or quenching of star formation in a
galaxy can alter its colour. However, such changes require a much
longer time scale. The effects of the tidal interactions in different
environments can be captured more reliably if we use both SFR and
colour for such studies.

The filaments are known to be somewhat denser region than the sheets.
We also study the SFR and colour of the major pairs in environments
with different local density and compare these finding to that
observed for the different geometric environments.

We organize the paper as follows: we describe the data and the method
of analysis in Section 2 and present the results and conclusions in
Section 3.

\section{Data and Method of Analysis}           

\subsection{SDSS Data}
\label{sec:data}
The Sloan Digital Sky Survey (SDSS) \citep{stout02} is currently the
largest redshift survey. It uses a dedicated 2.5 m telescope at Apache
Point Observatory in New Mexico to measure the spectra and images of
millions of galaxies in five different bands over roughly one third of
the sky. We download the SDSS data from the sixteenth data release of
Sloan Digital Sky Survey (SDSS) \citep{ahumada20} that are publicly
available at SDSS
Skyserver \footnote{https://skyserver.sdss.org/casjobs/}. We obtain
the spectroscopic and photometric information of all the galaxies
present within the region $135^{\circ} \leq \alpha \leq 225^{\circ}$
and $0^{\circ} \leq \delta \leq 60^{\circ}$. The spectroscopic and
photometric information of the galaxies are obtained from the
$SpecPhotoAll$ table.  We use $stellarMassFSPSGranWideNoDust$
\citep{conroy09} table to extract stellar mass and the star formation
rate of the galaxies. These estimates are based on the Flexible
Stellar Population Synthesis Models. The information of internal
reddening $E(B-V)$ for each galaxy is taken from $emissionlinesport$
table. The internal reddening are derived using the publicly available
Gas and Absorption Line Fitting (GANDALF) \citep{gandalf} and
Penalised PIXEL Fitting (pPXF) \citep{ppxf}. We set the
$scienceprimary=1$ while downloading our data to ensure that only the
galaxies with best spectroscopic information are included in our
analysis.

We find that the above mentioned properties are available for a total
350536 galaxies within the specified region. We restrict the $r$ band
apparent magnitude to $m_r\leq 17.77$ and construct three volume
limited samples with r-band absolute magnitude range $M_r \leq -19$,
$M_r \leq -20$, $M_r \leq -21$ that correspond to redshift limits
$z<0.0422$, $z<0.0752$ and $z<0.1137$ respectively. The total number
of galaxies present in the three volume limited samples corresponding
to $M_r \leq -19$, $M_r \leq -20$, $M_r \leq -21$ are 21984, 69456 and
85745 respectively.

We separately identify all the galaxy pairs in our data by employing
simultaneous cuts on the projected separation and the rest frame
velocity difference. Any two galaxies with $r_p<$ 150 kpc and $\Delta
v <$ 300 km/s are identified as a galaxy pair. A galaxy may appear in
multiple pairs provided these conditions are satisfied. We allow this
following \citet{scudder12b} who showed that excluding the galaxies
with multiple companion does not make any difference to their
results. These cuts yield a total 24756 galaxy pairs present within
the specific region of the sky chosen in our analysis.

We cross match the $SpecObjID$ of the galaxies in the volume limited
samples to that with the sample of identified galaxy pairs. The
cross-matching respectively provides us with 2581, 5441 and 3039
galaxy pairs in the three volume limited samples corresponding to $M_r
\leq -19$, $M_r \leq -20$ and $M_r \leq -21$. We employ a further cut
$1 \leq \frac{M_1}{M_2} \leq 10$ in the stellar mass ratio of the
galaxy pairs. This reduces the number of available galaxy pairs to
2024, 5014 and 3002 in the three volume limited samples.

A significant number of close galaxy pairs can not be observed simply
due to the finite aperture of the SDSS fibres. The spectra of two
galaxies within $55^{\prime\prime}$ cannot be acquired simultaneously
\citep{strauss02} which leads to under selection of galaxy pairs with
angular separation closer than $55^{\prime\prime}$. We compensate this
incompleteness effect by randomly culling $67.5\%$ of galaxies in
pairs having angular separation $>55^{\prime\prime}$ \citep{patton08,
  ellison08, patton11, scudder12b}.

After the culling, we are left with 737, 2203 and 1600 galaxy pairs in
the three volume limited samples. We then identify only the major
pairs in our samples by restricting the stellar mass ratio to $1 \leq
\frac{M_1}{M_2} < 3$. Finally, in the three volume limited samples, we
have 387, 1409 and 1255 major galaxy pairs that are formed by 739, 2672
and 2432 galaxies respectively.

We use a $\Lambda$CDM cosmological model with $\Omega_{m0}=0.315$,
$\Omega_{\Lambda 0}=0.685$ and $h=0.674$ \citep{planck18} for our
analysis.

\begin{table}
\centering
\begin{tabular}{|c|c|}
\hline
Local dimension & Geometric environment\\
\hline
$0.75 \leq D < 1.25$ & $D1$\\
$1.25 \leq D < 1.75$ & $D1.5$\\
$1.75 \leq D < 2.25$ & $D2$\\
$2.25 \leq D < 2.75$ & $D2.5$\\
$D \geq 2.75$ & $D3$\\
\hline 

\end{tabular}
\caption{This table shows range of local dimension values $D$ and the
  associated geometric environment of galaxies.}
\label{tabld}
\end{table}

\subsection{Morphology of the local environment}
\label{sec:ldim}
The galaxies reside in various types of geometric environment of the
cosmic web. We calculate the local dimension \citep{sarkar09} of each
galaxy to quantify the morphology of its local environment. The local
dimension of a galaxy is estimated from the number counts of galaxies
within a sphere of radius $R$ centered on it. The number counts of
galaxies within a given radius $R$ can be written as,
\begin{equation}
N(< R) = A\,R^D
\label{ld}
\end{equation}
where $A$ is a proportionality constant and $D$ is the local
dimension. For each galaxy, the radius of the sphere is varied over
length scales $R_1 \mpc \leq R \leq R_2 \mpc$. We consider only
those galaxies for which there are at least 10 galaxies available
within the two concentric spheres of radius $R_1$ and $R_2$. The
measured number counts $N(<R)$ within $R_1$ and $R_2$ are fitted to
\autoref{ld} and the best fit values of A and $D$ are determined using
a least-square fitting. We further estimate the goodness of each fit
by measuring the associated $\chi^2$ per degrees of freedom. Only the
fits with chi-square per degree of freedom $\frac{\chi^2}{\nu} \leq
0.5$ are considered for our analysis \citep{sarkar19}. We set $R_1=2
\mpc$ and $R_2=10 \mpc$ for the present analysis. The local
dimension $D$ characterizes the geometric environment around a galaxy.
A finite range of local dimension is assigned to each type of
morphological environment (\autoref{tabld}). We classify the
morphology of the surrounding environment of a galaxy based on these
definition. The $D1$-type galaxies reside in one dimensional straight
filament. A $D2$-type galaxy is embedded in a two-dimensional
sheet-like environment and $D3$-type galaxies are expected to be
surrounded by a homogeneous distribution in three-dimension. Besides,
there can be intermediate local dimension values that may arise when
the measuring sphere includes galaxies from multiple morphological
environments. For instance, $D1.5$-type represents an intermediate
environment between filaments and sheets.

\subsection{Local density of environment}
We estimate the local density of the environment of each galaxy using
the distance to the $k^{th}$ nearest neighbour in three-dimension. The
local density $\eta_k$ \citep{casertano85} around a galaxy is defined
as,
\begin{equation}
\eta_k = \frac{k-1}{V(r_k)}
\end{equation}
where $r_k$ is the distance to the $k^{th}$ nearest neighbour and
$V(r_k)=\frac{4}{3}\pi r^3_k$ is the volume of the sphere associated
with radius $r_k$. We set $k=5$ and consider the $5^{th}$ nearest
neighbour from each galaxy to compute the local density around it.
The local density would be underestimated near the boundary of the
survey volume. We also estimate the closest distance to the survey
boundary $r_b$ from each galaxy and compare it with $r_k$. We consider
only those galaxies in our analysis for which $r_k<r_b$. This discards
all the galaxies near the survey boundary.

We determine the median local density of each samples of major
pairs. Each sample is then divided into two subsamples based on its
median density. We consider the pairs to be hosted in the high density
regions if their local density lies above the median. Similarly the
pairs in the low density regions are defined as those having a local
density below the median value.

\begin{figure*}
\resizebox{14.4cm}{10cm}{\rotatebox{0}{\includegraphics{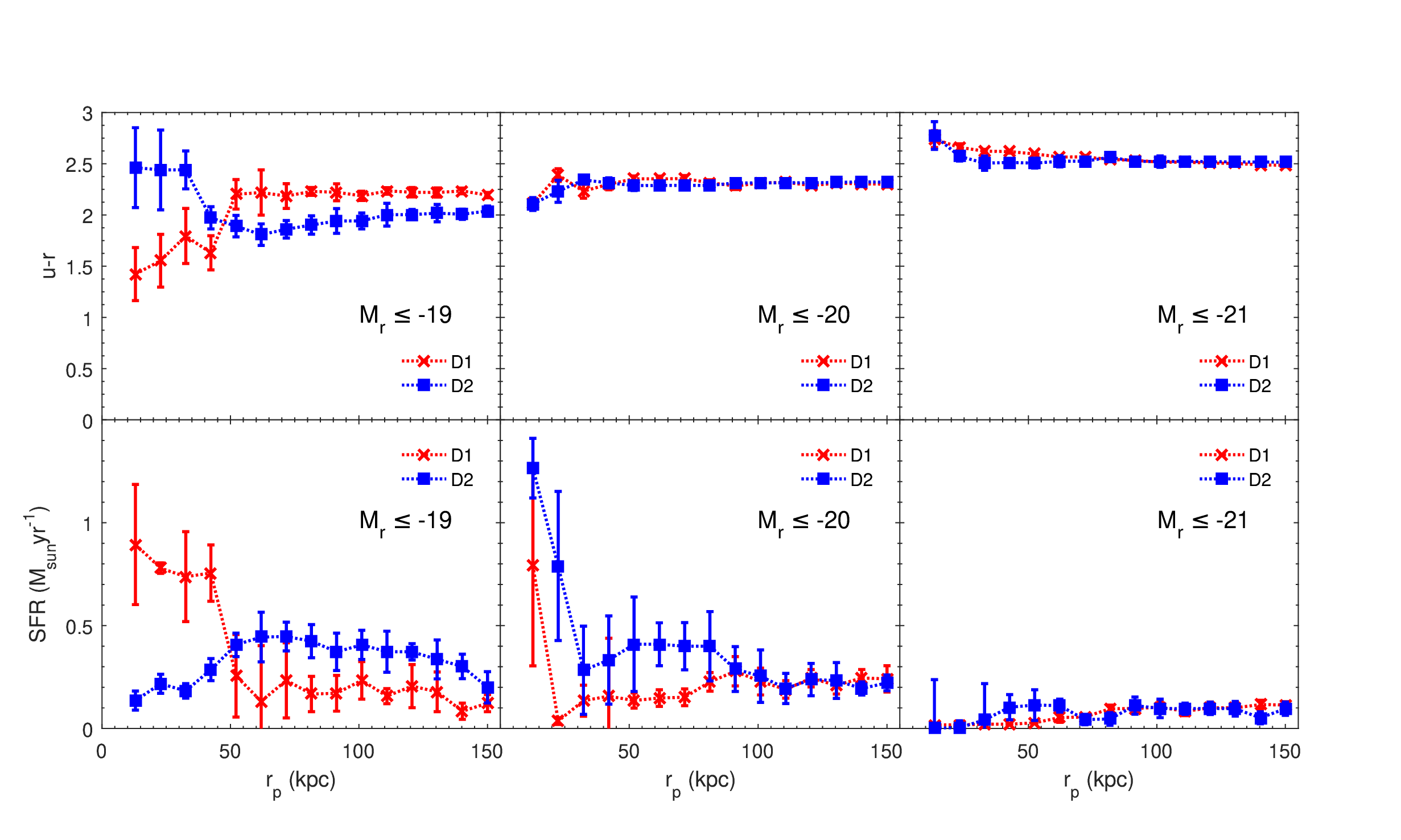}}}
\caption{The top left, top middle and top right panels show the
  cumulative median colour of the major pairs as a function of the
  projected separation for the three magnitude bins $M_r \leq -19$,
  $M_r \leq -20$ and $M_r \leq -21$ respectively. The bottom three
  panels show the cumulative median SFR of the major pairs in the
  three magnitude bins. We compare the results for the major pairs
  residing in sheets and filaments in each panel of this figure. The
  $1\sigma$ error bars at each data point are obtained from 10
  Jackknife samples drawn from each dataset.}
\label{Fig1}
\end{figure*}

\begin{figure*}
\resizebox{14.4cm}{10cm}{\rotatebox{0}{\includegraphics{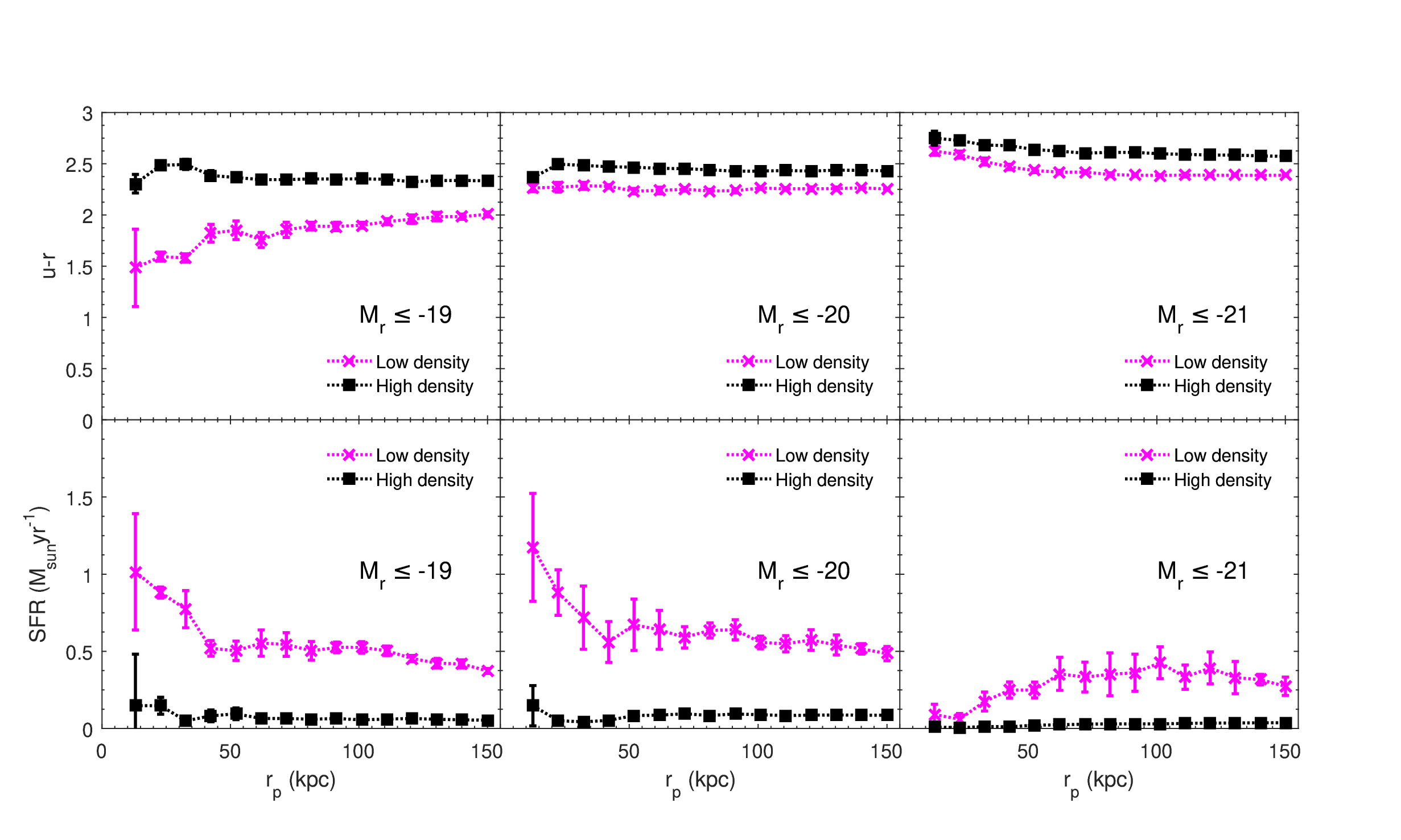}}}
\caption{Same as Figure~\ref{Fig1} but for the major pairs residing in
  the low-density and high-density regions.}
\label{Fig2}
\end{figure*}

\begin{figure*}
\resizebox{14.4cm}{10cm}{\rotatebox{0}{\includegraphics{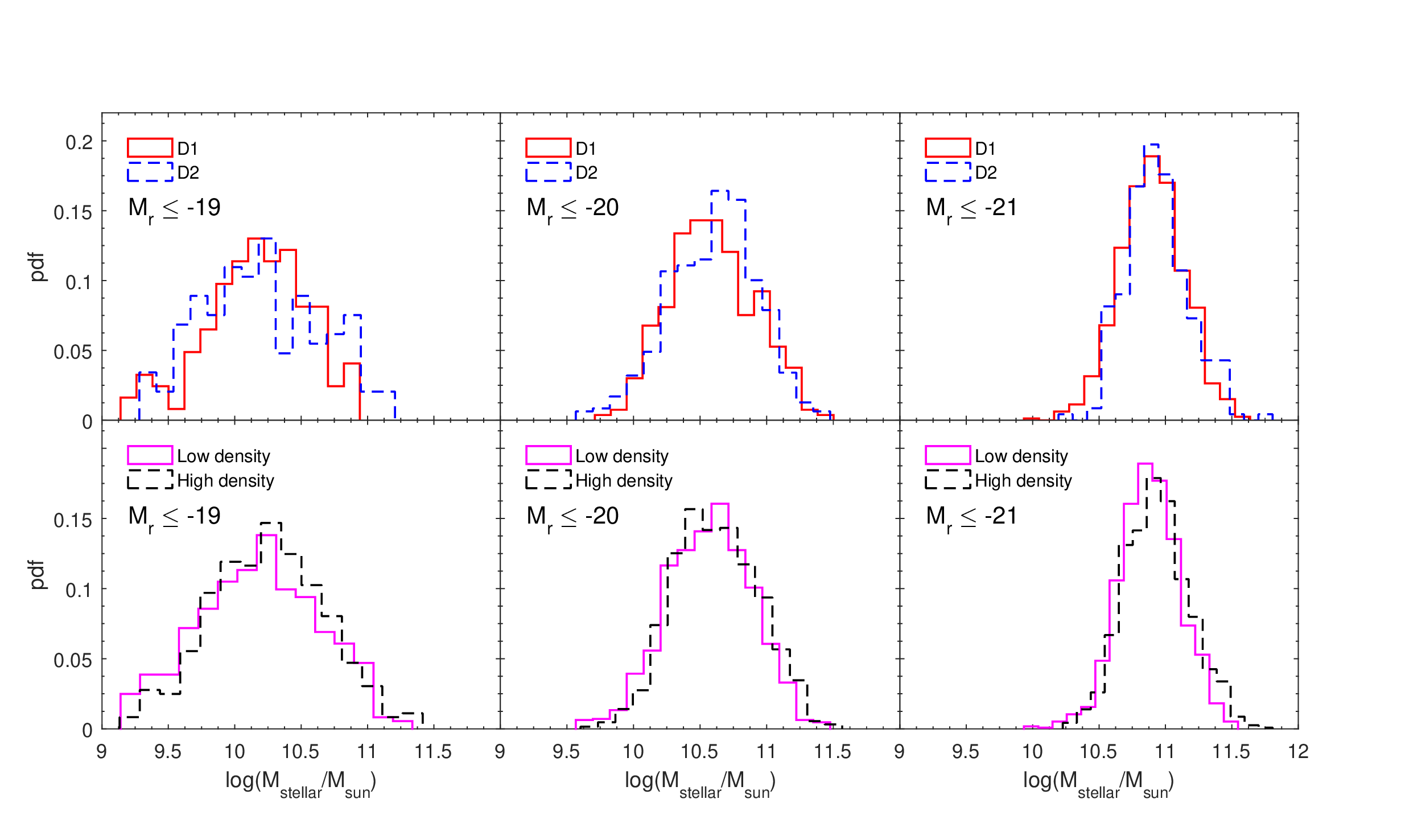}}}
\caption{The top three panels show the distributions of
  $log(M_{stellar}/M_{sun})$ for the major pairs residing in $D1$ and
  $D2$-type environment in the three volume limited samples. The three
  bottom panels compares the same but for the major pairs residing in
  the high-density and low-density regions.}
\label{Fig3}
\end{figure*}

\section{Results and Conclusions}
\label{sec:rescon}

We show the cumulative median of the dust corrected $(u-r)$ colour for
the major pairs as a function of the projected separation in sheets
and filaments in the top left panel of \autoref{Fig1}. The results in
this panel shows that at smaller pair separation, the major galaxy
pairs in the sheet-like structures are significantly redder compared
to those residing in the filamentary environments. We find a crossover
between the two curves at $\sim 50$ kpc beyond which the major pairs
in filaments are redder than those embedded in the sheet-like
structures. We repeat our calculations for the SFR in the major pairs
in similar manner. The results are shown in the bottom left panel of
\autoref{Fig1}. We find that the major pairs with a projected
separation $<50$ kpc are more star forming in filaments compared to
those hosted in the sheet-like environments. Interestingly, we also
notice a reversal of this behaviour at $\sim 50$ kpc for SFR similar
to that observed for the dust corrected $(u-r)$ colour. Again, the
major pairs with a projected separation greater than 50 kpc are more
star forming in sheets compared to those in filaments. The colour and
star formation rate are strongly correlated due to the observed
bimodality \citep{strateva01, baldry04, pandey20a}. A similarity in
the results for colour and SFR are not surprising. However, the
presence of the crossover at nearly the same length scale for both the
properties is certainly interesting.

A number of earlier works find a statistically significant alignment
of the galaxy pairs with their host filaments. Using the SDSS data,
\citet{tempel15} find $\sim 25\%$ extra aligned pairs in filaments
compared to a random distribution. A similar analysis of SDSS galaxy
pairs in filaments by \citet{mesa18} confirms the alignment signal and
suggests a stronger alignment closer to the filament spine. Such
preferred alignment indicates an anisotropic accretion within the
filaments. The interactions between the galaxies in the aligned pairs
could be more effective in triggering new star formation. We propose
that the trends observed in the top left and bottom left panels of
\autoref{Fig1} may arise due to the preferred alignment of galaxy
pairs inside filaments.

We repeat our analysis for volume limited samples constructed in two
other magnitude bins. This would reveal any luminosity dependence of
these results. The results for the magnitude bins $M_r \leq -20$ and
$M_r \leq -21$ are respectively shown in the top/bottom middle and
top/bottom right panels of \autoref{Fig1}. Interestingly, the trends
observed in the magnitude bin $M_r \leq -19$ are not present in the
brighter samples. The galaxy pairs in the filaments and sheets from
the brighter galaxy samples trace the higher density regions in these
structures. The star formation of galaxies are known to be suppressed
in the high-density regions. The red galaxies usually have
$(u-r)>2.22$ \citep{strateva01}. It is interesting to note that the
cumulative median colour of the major pairs in the brighter samples
are greater than $2.22$ at nearly all pair separation. This clearly
indicates that the major pairs in the high density regions of the
filaments and sheets are not effective in forming new stars. Both the
local density and the large-scale environment are important in the
formation and evolution of galaxies. But the local density is known to
play a more dominant role. The absence of these trends in the brighter
samples possibly indicates the dominance of the local density over the
large-scale environment.

We separately study the effects of the local density in deciding the
colour and SFR of the interacting major pairs. We split each samples
of major pairs into two based on their median density. This provides
us two sets of major pairs corresponding to low and high density
regions. The results of this analysis are shown in \autoref{Fig2}. The
top/bottom left, top/bottom middle and top/bottom right panels of
\autoref{Fig2} respectively show the results corresponding to
magnitude bins $M_r \leq -19$, $M_r \leq -20$ and $M_r \leq -21$. The
results are qualitatively similar in the three magnitude bins. We note
that at each pair separation, the cumulative median of the dust
corrected $(u-r)$ colour and SFR of the major pairs are different in
the low-density and high-density regions. The major pairs in the low
density regions are more star forming and bluer as compared to their
high-density counterparts. The differences in colour and SFR decrease
with the increasing pair separation but no crossover is observed
between the curves in any of the volume limited samples. The
differences in colour and SFR persist at each projected pair
separation upto 150 kpc for all three volume limited samples. This
indicates that the local density and large-scale environments affect
the galaxy interactions in noticeably different manner. We also note
that the differences between the colour and SFR at each pair
separation are significantly smaller for the brighter samples. The
pairs in the brighter samples preferentially inhabit the denser
regions. Consequently, the pairs in these samples have smaller
differences in their local density.

%%%%%%%%%%%%%%%%%%%%%%%%%%%%%%%%%%%%%%%%%%
\begin{table*}
\centering
\begin{tabular}{|c|c|c|c|c|c|c|c|}
Magnitude bin & Major pairs in & $D_{KS}$ & \multicolumn{5}{c}{$D_{KS}(\alpha)$}\\
\hline
& & & 99\% & 90\% & 80\% & 70\% & 60\% \\
& D1, D2 type & 0.1323 & 0.1992 & 0.1498 & 0.1313 & 0.1192 & 0.1098 \\
$M_r \leq -19$ & Low density, High density & 0.0743 & 0.1211 & 0.0910 & 0.0798 & 0.0724 & 0.0667 \\
\hline
& D1, D2 type & 0.0735 & 0.1031 & 0.0776 & 0.0680 & 0.0617 & 0.0568\\
$M_r \leq -20$ & Low density, High density & 0.0688 & 0.0646 & 0.0486 & 0.0426 & 0.0386 & 0.0356\\
\hline
& D1, D2 type & 0.0747 & 0.1213 & 0.0912 & 0.0799 & 0.0726 & 0.0668\\
$M_r \leq -21$ & Low density, High density & 0.0899 & 0.0678 & 0.0510 & 0.0477 & 0.0406 & 0.0374\\
\hline

\end{tabular}
\caption{The above table shows Kolmogorov-Smirnov statistic $D_{KS}$
  for comparison of $log(M_{stellar}/M_{sun})$ of major pairs residing
  in $D1$, $D2$ type environment and low density, high density
  regions. This table also shows the critical values $D_{KS}(\alpha)$
  above which null hypothesis can be rejected at different confidence
  levels}
\label{tab:ks}
\end{table*}
%%%%%%%%%%%%%%%%%%%%%%%%%%%%%%%%

It is well known that the colour and SFR of galaxies are strongly
correlated with the stellar mass. So the observed differences in the
properties of interacting galaxies in different environments may also
arise due to a difference in their stellar mass. We investigate this
possibility by performing Kolmogorov-Smirnov (KS) test on the stellar
mass distributions of the galaxy pairs in different environments. We
compare the probability distribution function of the stellar mass for
the major pairs residing in $D1$ and $D2$-type environment in the top
three panels of \autoref{Fig3}. We carry out a similar comparison for
the pairs in low and high-density regions in the three bottom panels
of \autoref{Fig3}. The results of the KS tests are summarized in
\autoref{tab:ks}. We find that the stellar mass distributions of the
interacting galaxy pairs in $D1$ and $D2$-type environments are not
significantly different. The null-hypothesis can not be rejected at
very high confidence level for all the three volume limited
samples. So the observed differences in the colour and SFR of
interacting galaxies in filaments and sheets do not originate from the
differences in their stellar mass. However, the results of the KS test
suggest that the stellar mass distributions of the galaxy pairs in the
low-density and high-density regions are significantly different for
the last two magnitude bins. So the stellar mass may have a role in
causing the differences in the properties of the interacting galaxies
in the low-density and high-density regions.

Generally, filaments are denser than sheets. So one would expect that
the interacting galaxy pairs in filaments to be less star forming and
redder than those residing in sheets. However we observe an exactly
opposite trend in our analysis for the galaxy pairs with projected
separation less than 50 kpc. This indicates that the local density and
large-scale environments affect the galaxy interactions in noticeably
different manner. The local density is known to play a more dominant
role. The absence of the effects of large-scale environments in the
brightest sample in our analysis possibly indicates the dominance of
the local density over the large-scale environment. It is worth
mentioning here that the effects of local-density and large-scale
environment are coupled with each other. One may study the impact of
the large-scale environment by conditioning the local environment and
vice versa. However this drastically reduces the number of pairs
available for this study. Another limitation of this study is that the
three magnitude bins used here are not completely independent. This
introduces some ambiguity in the interpretation of our results. We
find that the use of the independent magnitude bins also drastically
reduces the number of available pairs.

Our study clearly shows that the colour and SFR in the interacting
galaxies are not only affected by the local density but also by their
large-scale morphological environment. We note that the effects of the
local density and the morphological environment are quite distinct
from each other. We conclude that the large-scale structures such as
filaments and sheets play a fundamental role on the outcomes of galaxy
interactions. The present analysis only classifies the pairs based on
their local density and local dimension. It would be interesting to
carry out a similar analysis with a set of individual sheets and
filaments.  We plan to carry out such an analysis in a future
work. This would help us to understand better the effects of alignment
on galaxy interactions in filaments and sheets.

\begin{acknowledgements}
The authors thank the SDSS team for making the data publicly
available. BP would like to acknowledge financial support from the
SERB, DST, Government of India through the project CRG/2019/001110. BP
would also like to acknowledge IUCAA, Pune for providing support
through associateship programme. SS acknowledges IISER Tirupati for
support through a postdoctoral fellowship.

Funding for the SDSS and SDSS-II has been provided by the Alfred
P. Sloan Foundation, the Participating Institutions, the National
Science Foundation, the U.S. Department of Energy, the National
Aeronautics and Space Administration, the Japanese Monbukagakusho, the
Max Planck Society, and the Higher Education Funding Council for
England. The SDSS website is http://www.sdss.org/.

The SDSS is managed by the Astrophysical Research Consortium for the
Participating Institutions. The Participating Institutions are the
American Museum of Natural History, Astrophysical Institute Potsdam,
University of Basel, University of Cambridge, Case Western Reserve
University, University of Chicago, Drexel University, Fermilab, the
Institute for Advanced Study, the Japan Participation Group, Johns
Hopkins University, the Joint Institute for Nuclear Astrophysics, the
Kavli Institute for Particle Astrophysics and Cosmology, the Korean
Scientist Group, the Chinese Academy of Sciences (LAMOST), Los Alamos
National Laboratory, the Max-Planck-Institute for Astronomy (MPIA),
the Max-Planck-Institute for Astrophysics (MPA), New Mexico State
University, Ohio State University, University of Pittsburgh,
University of Portsmouth, Princeton University, the United States
Naval Observatory, and the University of Washington.
\end{acknowledgements}

\label{lastpage}
\end{document}